\documentclass[twocolumn]{article}
\usepackage{graphicx}
\usepackage{amsmath} 
\usepackage{graphicx}
\usepackage{subcaption}
\usepackage{booktabs}

\usepackage{amsmath}
\usepackage{amssymb}
\usepackage{amsthm}

\usepackage{algpseudocode}
\usepackage{algorithm}

\usepackage[a4paper, total={6.5in, 9in}]{geometry}

\title{Neural Network Approach to Demand Estimation and Dynamic Pricing in Retail}
\author{Kirill Safonov}
\date{\today}

\begin{document}

\twocolumn[
  \begin{@twocolumnfalse}
    \maketitle
    \begin{center}
      \begin{minipage}{0.77\textwidth} 
        \begin{abstract}
          \noindent This paper contributes to the literature on parametric demand estimation by using deep learning to model consumer preferences. Traditional econometric methods often struggle with limited within-product price variation, a challenge addressed by the proposed neural network approach. The proposed method estimates the functional form of the demand and demonstrates higher performance in both simulations and empirical applications. Notably, under low price variation, the machine learning model outperforms econometric approaches, reducing the mean squared error of initial price parameter estimates by nearly threefold. In empirical setting, the ML model consistently predicts a negative relationship between demand and price in 100\% of cases, whereas the econometric approach fails to do so in 20\% of cases. The suggested model incorporates a wide range of product characteristics, as well as prices of other products and competitors.
          \vspace{2em}
        \end{abstract}
      \end{minipage}
    \end{center}
  \end{@twocolumnfalse}
]

\section{Introduction}

Demand estimation and determining the optimal price are one of the key challenges in company management. Retail companies face an even more complex task, as they simultaneously sell thousands of products where the assortment and the competitive environment are constantly changing.
In the e-commerce setting, the need for dynamic pricing and accurate demand estimation is especially relevant due to the competitive nature of online marketplaces. 

 Traditional econometric models for demand estimation often rely on significant price variation to accurately capture consumer responsiveness. However, in many real-world applications, price variation may be quite low due to pricing regulations or particular pricing policies.

In this work, I propose a novel approach to demand estimation that uses a neural network to predict the functional form of demand and compare it with well-established econometric model (such as Pinkse et al., 2002). The machine learning (ML) model is designed to overcome the challenges that come with low price variation by extracting latent patterns from item-specific characteristics and environmental variables\footnote{The replication package is avaiable at https://github.com/krsafonov/pricing}.  Through both simulation and empirical analysis, I demonstrate that the ML model performs ``well'' even in the environment of low price variation, where the econometric model struggles. This finding suggests that machine learning techniques offer a robust alternative for demand estimation in retail contexts. 

The revenue management literature studies dynamic pricing from the perspective of maximizing revenue by adjusting prices based on demand fluctuations, inventory levels, and market conditions, see Talluri \& van Ryzin (2006), Özer \& Phillips (2012), and den Boer (2015) for comprehensive reviews of the literature. 
Earlier research in the literature assumes that the demand follows an exogenous distribution that is known to the decision makers (Gallego \& van Ryzin, 1994), whereas more recent literature considers the problem of dynamic pricing where demand function is unknown a priori which is a more realistic assumption. Researchers address the demand uncertainty through both parametric (e.g. Dimitris \& Perakis, 2006, Farias \& van Roy, 2010, Harrison et al., 2012) and non-parametric approaches (e.g. Besbes \& Zeevi, 2009, Wang et al., 2014). This work falls in the stream of literature that adopts a parametric approach. 

This paper contributes to the literature by estimating demand in settings with a large number of SKUs, minimal price variation for individual products, while also accounting for competitors' prices in real time. The overall optimization framework is similar to Fisher et al. (2018), however, I do not setup the experiment to estimate the demand. In prior studues like Nevo (2001) and  Pinkse
et al. (2002), a common approach to estimating product-level demand functions is defining a panel econometric model. However, Fisher et al. (2018) highlight the challenges in this process, particularly due to limited price variation within a single product, emphasizing that these difficulties persist when using Hausman-type instruments or lagged prices.

As Aparicio et al. (2022) note that online retailers are much more prone to frequent price changes than their offline competitors, which creates a great demand for effective pricing models. Meanwhile application of deep learning methods remains understudies in the retail pricing literature. Liu et al. (2021) implement a reinforcement learning algorithm for dynamic pricing and validate it through a field experiment. However, to the best of my knowledge, there are very few examples of applying classical deep learning methods to this problem. This article eliminates this gap and compares machine learning models with classical econometric ones.

The rest of the paper is structured as follows: in Section ~\ref{sec:framework}, I outline the theoretical framework of the machine learning approach and detail the econometric model used for comparison. Section ~\ref{sec:simulation} describes the simulation setup and presents the results that illustrate the performance of both models under varying conditions of price variation. In Section ~\ref{sec:data}, I discuss the data used for the empirical analysis, data preparation and feature engineering techniques. Section ~\ref{sec:empiricalresult} presents the empirical results and Section ~\ref{sec:conclusion} concludes.

\section{Framework} \label{sec:framework}
\subsection{Machine Learning Approach}
    The machine learning approach for demand estimation uses a neural network to extract relevant information where the main objective is to predict the parameters of the structural demand model. The results of this method should generalize well, particularly in scenarios where historical data lacks sufficient variation in prices. Formally, we aim to estimate the parameters $\theta_i$ of the heterogeneous demand model $\mathcal{D}_i(P_i, X_i, S_i)$, where $P_i$ denotes the price of item $i$, $X_i$ represents the environmental variables, and $S_i$ represents item-specific characteristics. Importantly, $X_i$ is not included directly in the demand function but rather determines the parameters $\theta_i$.

    Primarily, the network $\mathcal{F}_{\text{emb}}(S_i)$ estimates the vector representation of item $i$ in a latent feature space $Z$. Note that when the model is trained, $Z_i$ remains constant for the item $i$. This approach estimates embeddings as part of a joint optimization problem, which differs from methods that rely on pre-evaluated embeddings. The resulting embedding is then combined with the environmental variables and used as input to $\mathcal{F}_{\text{FC}}(Z_i, X_i)$, which predicts $\theta_i$ for the structural model. Although embeddings may not have a direct practical interpretation in structuring the network, they provide a convenient way to reduce the dimensionality of input parameters. Additionally, the output of this internal model $\mathcal{F}_{\text{emb}}$ can serve as embeddings for other tasks. This approach of using deep learning to estimate structural models is similar to the framework developed by Farrell et al. (2020). In summary, the model is defined as:
    
    \begin{equation}
    \theta_i = \mathcal{F}_{\text{FC}}(\mathcal{F}_{\text{emb}}(S_i), X_i) \label{eq:general_network}
    \end{equation}

    The loss is estimated as $\mathcal{L}(q_i, \mathcal{D}_{\theta_i}(P_i))$, where $q_i$ represents the number of sales of good $i$. Notably, the neural network does not directly observe the price $P_i$, which helps to create a more generalized model by avoiding an overemphasis on specific focal prices.

\subsection{Econometric Approach}
    An alternative approach to extracting price elasticity is to apply more conventional econometric models. Here, I examine the case where competition among firms is spatial and determined by the topology of the product space. Following the approach of Pinkse et al. (2002), I project product characteristics within this space onto a flexible function. This method is computationally efficient and allows to estimate the parameters with OLS:
    \begin{equation}
    \ln{q_{jt}} = \alpha_j + \beta_j\ln{p_{jt}} + f(\bar d_j, \gamma)^\top \bar p_t + \epsilon_{jt}\label{eq:econ_model}
    \end{equation}
    This models includes the prices $p_{jt}$, the distance vector $\bar d_j$, which represents the distances from each product $j$ to all other products (including itself), and the price vector $\bar p_j$, which captures the prices of other products within the same category at time $t$. The functional form of $f$ is open to discussion; however, I adopt a polynomial form ~\eqref{eq:f} of degree 3. The error term is normally distributed.
    \begin{equation}
    f(d, \gamma) = \sum_{i=0}^n \gamma_i d^i \label{eq:f}
    \end{equation}
    
    This model faces several challenges, including price endogeneity and interdependence arising from demand shocks that affect the entire category simultaneously. However, I evaluate the proposed model with a strong assumption of price shocks independence, while other approaches, such as BLP models (Berry, Levinsohn, and Pakes, 1995) or experiments with price randomization (Fisher et al., 2018), could help address these issues\footnote{See Conlon \& Gortmaker (2020) for recent methodological advances of BLP-type problems.}. Finally, the biggest challenge is not necessarily these issues, but rather the small price variation.

    Formally, it can be demonstrated that the variance of the price coefficient estimator is inversely related to the variance of the price of the good. Moreover, the more prices in the market are correlated (collinearity of price movements is measured in the $R^2$ of the regression $p$ on $f(\bar d_j, \gamma)^\top \bar p_t$), the more the variance of the price coefficient is. It gives us theoretical grounds on the relationship between the mean squared error of the estimated coefficient and the variance of the price:
    \begin{equation}
        \text{MSE}(\hat{\theta}) = \mathbb{E}[(\hat{\theta} - \theta)^2] = \text{Var}(\hat{\theta}) + \text{Bias}(\hat{\theta})^2 \label{eq:mse}
    \end{equation}
    $$
    \text{MSE}(\hat{\theta}) \propto \frac{1}{\text{Var}(p)}
    $$
    $$
    \text{MSE}(\hat{\theta}) \propto R^2
    $$

    \noindent \textbf{Lemma.} \textit{The variance of the price coefficient is inversely proportional to the variance of the price.}
    \begin{proof}
        The variance of the estimator $\hat{\beta}$ is $\text{Var}(\hat{\beta}) = \left[ (\mathbf{X}^T \mathbf{X})^{-1} \right]_{22} \sigma^2$, where $\mathbf{X}$ is the matrix of independent variables of \eqref{eq:econ_model}.

        According to Frisch–Waugh–Lovell theorem, $\hat \beta$ can be estimated from the regression of $q$ on $\hat{p}_{\text{res}}$, where $\hat{p}_{\text{res}}$ is the residual of the regression $p$ on other covariates. Hence, $\text{Var}(\hat{\beta}) = \frac{\sigma^2}{\sum (\hat{p}_\text{res} - \bar{p}_\text{res})^2}$.\\\\ In turn, $\text{Var}(\hat{p}_{\text{res}}) = (1 - R^2) \cdot \text{Var}(p)$
        \begin{equation}
            \text{Var}(\hat{\beta}) = \frac{\sigma^2}{\sum_{i=1}^n (p_i - \bar{p})^2 \cdot (1 - R^2)}
        \end{equation}
        where $R^2$ refers to the regression of $p$ on other covariates. 
    \end{proof}
\subsection{Optimization}

In this section, I briefly discuss the implementation of the estimated elasticity results in pricing strategy. The ML model estimates the functional form of the demand function which can be directly used in the revenue optimization while the regression results can be used with linear approximation. During inference, the model predicts the demand for the next day relying on previous sales, the product charestristics, subsititutes' prices and competitors' actions. 

In practice, a retailer typically aims to maximize a proxy for profit or imposes constraints on revenue rather than directly maximizing revenue. I will describe the second case, where constraints are imposed on overall and product margins, as a non-linear optimization problem:
\begin{equation}
\max_{p_{1t}, p_{2t}, ..., p_{It}} = \sum_{i=1}^I p_{it} \mathcal{D}_{\theta_{it}}(p_{it}) \label{eq:optimization}
\end{equation}
$$
\text{s.t.}\quad \frac{\sum_{i=1}^I (p_{it} - c_{it}) \mathcal{D}_{\theta_{it}}(p_{it})}{\sum_{i=1}^I p_{it} \mathcal{D}_{\theta_{it}}(p_{it})} \geq \text{margin target},
$$
$$
\text{margin LB}_i \leq \frac{p_{it} - c_{it}}{p_{it}} \leq \text{margin UB}_i
$$
This non-linear problem is not concave in price, thus, an optimizer is not guaranteed to find the global maximum. Iterating over different random initial parameters can help to find the best value.

To implement this model in practice, one should first collect sufficient historical data for training. Then, retrain the demand estimation model weekly using the accumulated data to forecast future demand. Finally, use the model's predictions to optimize pricing by solving the corresponding optimization problem, setting prices based on forecasted demand and business objectives.

\section{Simulation} \label{sec:simulation}
\begin{figure}[!tbh]
    \centering
    \begin{subfigure}[b]{0.3\textwidth}
        \centering
        \includegraphics[width=\textwidth]{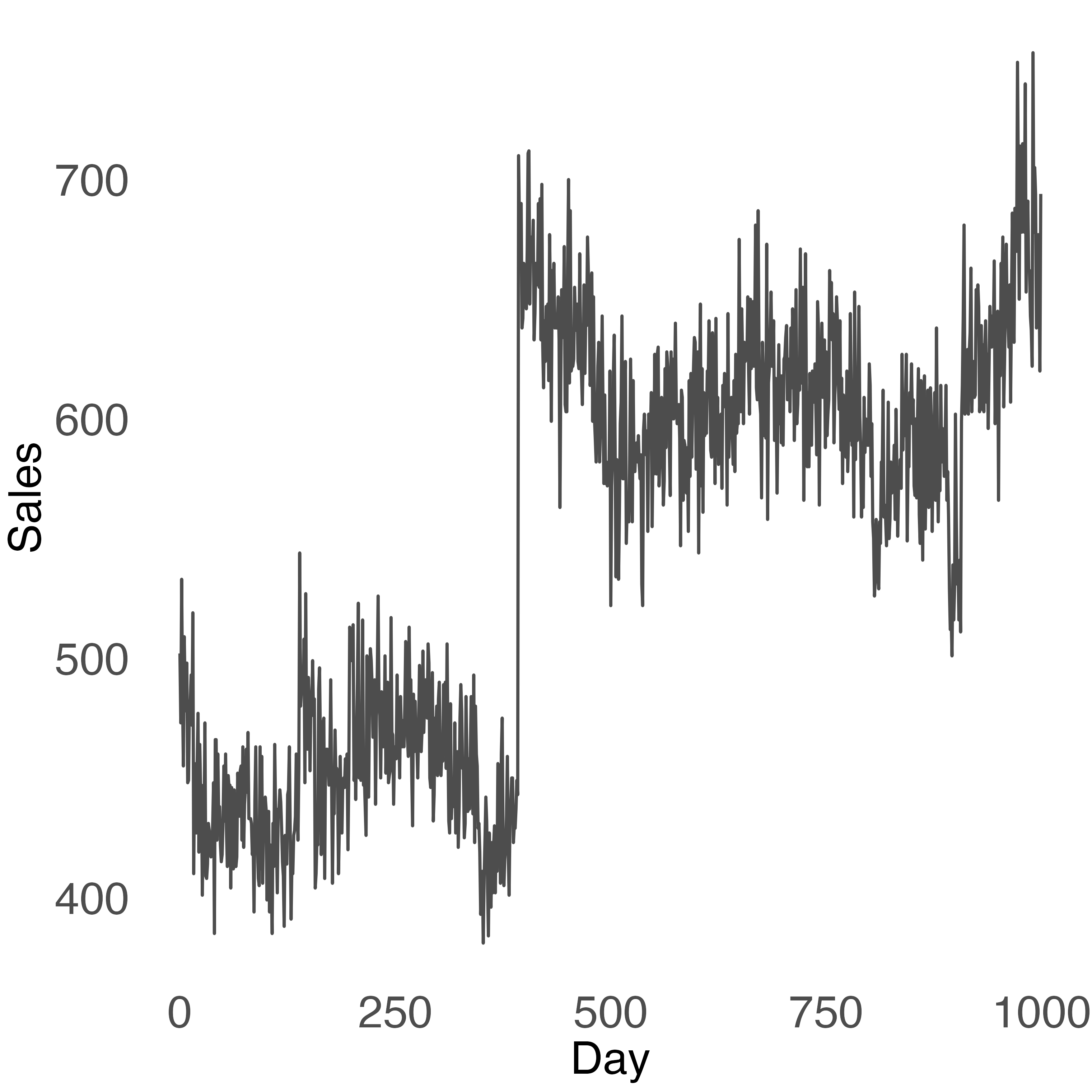}
        \caption{Sales over time}
        \label{fig:image1}
    \end{subfigure}
    \hfill
    \begin{subfigure}[b]{0.3\textwidth}
        \centering
        \includegraphics[width=\textwidth]{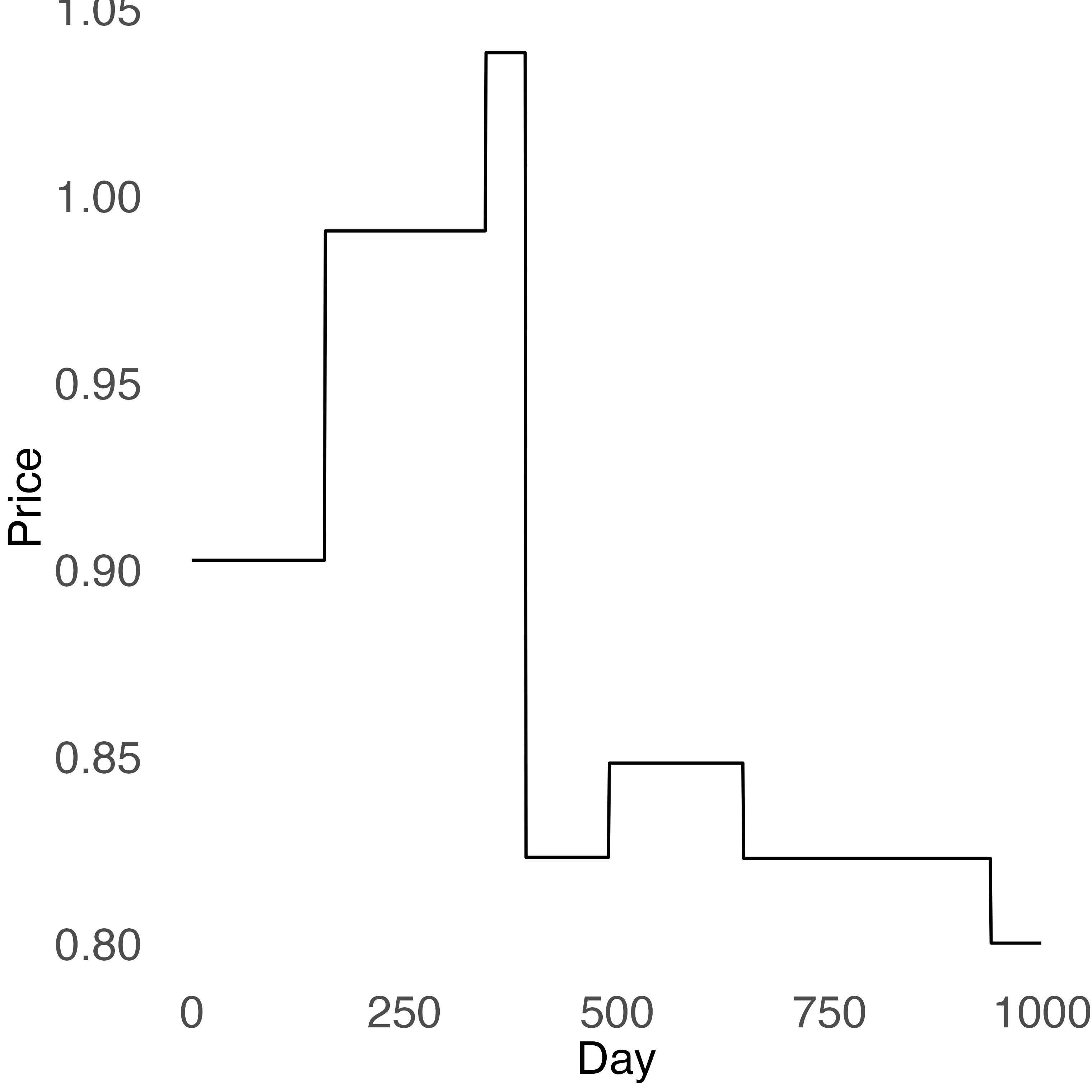}
        \caption{Price over time}
        \label{fig:image2}
    \end{subfigure}
    \caption{The example of the data generating process for one product with ($\varepsilon = 0.01$).}
    \label{fig:product_example}
\end{figure}

I create a category of $N$ products in which consumers choose goods based on product characteristics. First, there are significant $c^{\textit{sig}}$ and imaginary $c^{\textit{ima}}$ features. Significant features are those that consumers use to differentiate between goods and make purchasing decisions, while imaginary features are observed in the dataset but do not influence consumer choices (true coefficients for them are zero). By adding imaginary features, I introduce uncertainty into the product space, which mirrors real-world scenarios where some characteristics may be observed but do not actually affect consumer decisions. The true coefficients of the features are $\delta$. Let us set aside the process of obtaining these characteristics and assume they are already available, while remaining unaware of which characteristics are real and which are not.
Second, consumers pay attention to the price $p$ and each product $j$ has the elasticity parameter $\beta_j$ common to all consumers. Finally, the utilities $u_j$ ~\eqref{eq:utility} underlie the logit model and consumers make the random choice which product to buy according to probabilities $\pi_j$ ~\eqref{eq:logit}.
\begin{equation}
u_j = \beta_j p_j + \delta^\top c_j^{\textit{sig}} \label{eq:utility}
\end{equation}
\begin{equation}
\pi_j = \frac{\exp{u_j}}{\sum_{i=1}^N \exp{u_i}} \label{eq:logit}
\end{equation}
After generating product characteristics and setting the initial prices, the following steps occur. First, on each day, consumers decide which product to buy, thus, determining the sales. Second, with the probability $\varepsilon$, each product can change price by some small percentage (sampled from the normal distribution with zero mean). Overall, $T$ days are generated. 
In my example, I create 25 products with 10 characteristics but consumers pay attention only to 6 of them. The dataset includes 100,000 customers and spans 1,000 days. The real elasticity parameters are randomly generated and all are negative with the average of -2.75. See Figure ~\ref{fig:product_example} for an example of the data generating process.

\begin{figure}[!tbh]
    \centering
    \includegraphics[width=0.7\linewidth]{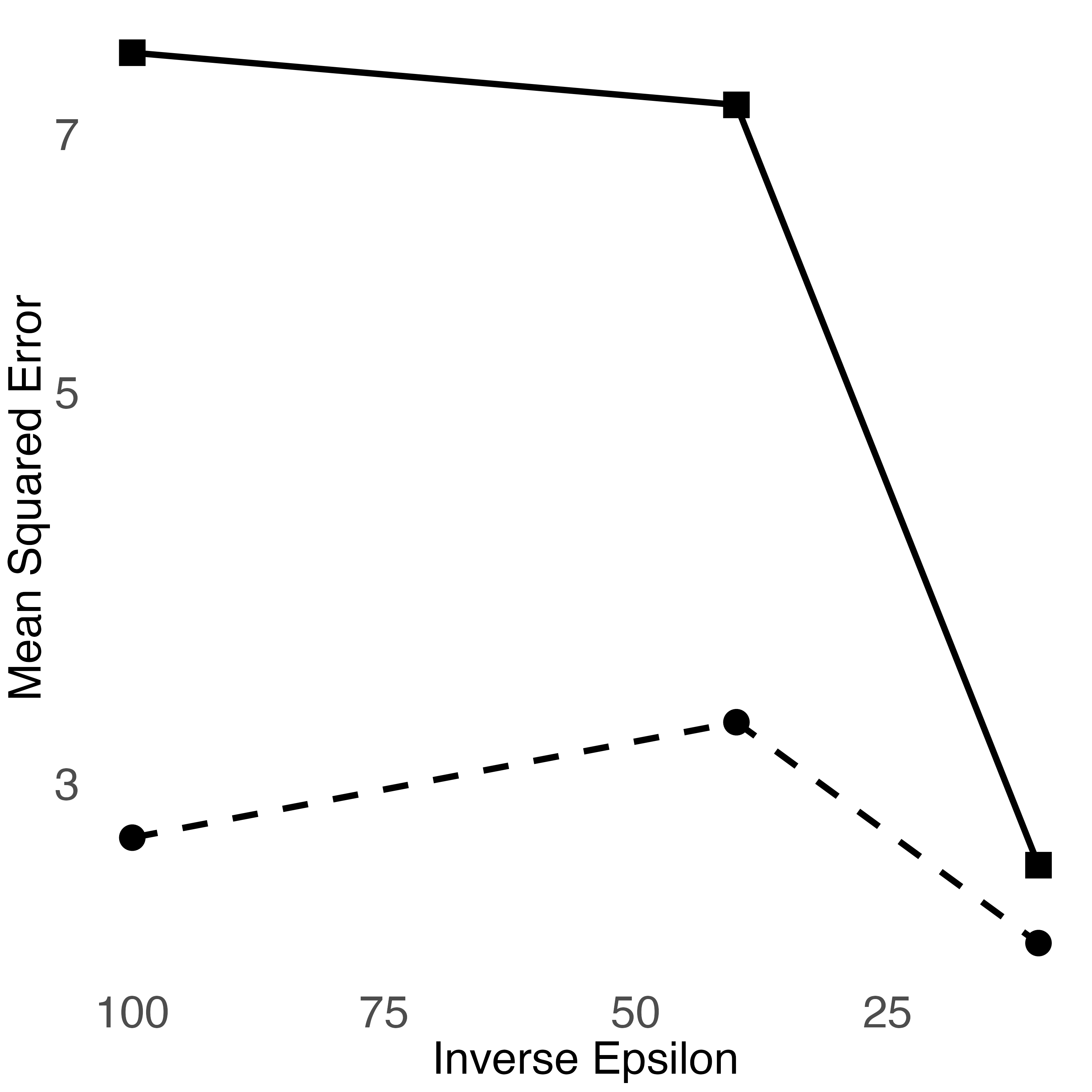}
    \caption{MSE comparison for different values of $\varepsilon$ (the x-axis represents inverse of $\varepsilon
    $). The dashed line is ML model results; the solid line represents the regression model results.}
    \label{fig:mse}
\end{figure}

I apply both models described above to 3 samples of generated data with $\varepsilon \in \{0.01, 0.025, 0.01\}$. The machine learning method directly learns from all features without any projections and functional forms assumptions required in ~\eqref{eq:econ_model}. The ML model shows reliable and stable results for all $\varepsilon$ values while the regression approach perfoms poorly when the price variation is low. Both models show strong results when the price changes are frequent ($\varepsilon = 0.1$).
See Figure ~\ref{fig:mse} for the models' performance comparison with different $\varepsilon$ values.

\section{Data} \label{sec:data}
    The data is comprised of a large e-commerce retailer's records on sales, prices, assortment availability collected daily over a span of 1.5 years. Additionally, I gather data on the price's of the major counterparts. Most of the time, the prices were the same in all branches (their number greatly exceeds 100) which makes it challenging to extract useful information. However, as part of the company's experiment, prices were slightly and uniformly increased in one-third of the stores, decreased in another third, and left unchanged in the remaining stores for a period of five months\footnote{This experminetal design raises concerns for two main reasons: first, all prices in the affected stores were increased simultaneously and by the same amount; second, it didn't try to use the potential for intra-store and inter-store randomization to mitigate the effect of common shocks. The results of this experiment, conducted by the company in the style of A/B testing, showed that 38\% of the elasticity estimates were positive, along with a large number of outliers.}.

    The primary numerical signals include the price itself, along with aggregated and average sales over a time window prior to the current date\footnote{The values are aggregated using metrics such as mode, median, weighted average, standard deviation, coefficient of variation, and an exponentially weighted average.}. Other variables are a mean value of on-shelf availability over the same time window, price of others goods in this category, and competitors' prices. Competitors' prices are aggregated in the same manner as the company's price and averaged across the available competitors since the set of competitors varies between products. It is important to use information aggregated over a period before the prediction since there is strong autocorrelation between today's price and yesterday's, and we want to avoid this spillover effect. Finally, we incorporate the date information by adding the day of the week and the week number, as there is significant seasonality and clear patterns in purchase behavior throughout the week.

    It is evident that there exists cross-elasticity between products. To address this, I use the company's product tree, which is a hierarchical structure that shows how products are related. The top level includes broad categories, such as dairy or pastry, while the most granular levels represent detailed classifications, such as dairy-free ice cream weighing less than 1/2 lbs. This tree does not address how any two goods are related from the customer's perspective but indicates that the goods share similar properties. I add the item's position in this tree as a series of levels (level1 value, level 2 value, ..., level 8 value). Level 1 consists of 20-30 different values, while level 8 contains thousands of values.

    \begin{figure}
        \centering
        \includegraphics[width=0.6\linewidth]{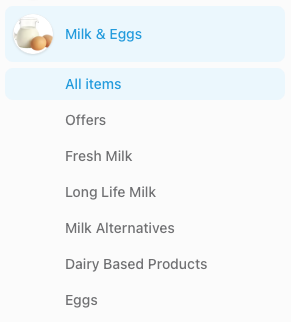}
        \caption{Example of a product tree: level 1 and level 2 for milk and eggs category.}
        \label{fig:tree-example}
    \end{figure}

    I always add the unique identifier of a good and its upper category in the tree. Other tree information is truncated up to level 4 and then passed through an embedding layer to obtain embeddings of size 64. Using deeper levels significantly increases the number of features, which can lead to instability or processing delays, making it challenging to manage within a reasonable amount of time and computational resources.

    I focus on the most valuable subsamples for my analysis since prices are usually stable. I only include observations where the price deviates by more than 5\% from the average, which reduces the dataset from over 10 million to about 3 million observations. Finally, most of the numerical features are normalized and transformed using a logarithmic scale.

\section{Empirical result} \label{sec:empiricalresult}
\subsection{Machine Learning Approach}
The network consists of two subsequent sub-networks. The first one ($\mathcal{F}_{\text{emb}}$) encodes the categorical data into the vector of size 64 (a product embedding). The embedding is then combined with the numerical features, and the second network ($\mathcal{F}_{\text{FC}}$) predicts the parameters of the demand function. The first network consists of three linear layers with dropout, while the second network has five layers with both dropout and batch normalization.

\begin{equation}
\mathcal{D}_{\theta_i}(P_i) = \alpha + \beta P_i \label{eq:demand_function}
\end{equation}

$$
\theta_i = 
\begin{pmatrix}
\alpha_i \\
\beta_i
\end{pmatrix}
$$

The equation ~\ref{eq:demand_function} is a simple linear demand function which was used in the estionation of the model. Other promising candidates are the parametrized functions such as $\mathcal{D}_{\theta_i}(P_i) = \sum^T_{t=0} \max{(a_t + b_t P_i, 0)}$ or any combination of non-linear functions. Hence, the network on each observation estimates only two parameters. The loss function is mean squared error: $\tfrac{1}{n} \sum^N_{i=0} (\mathcal{D}_{\theta_i}(P_i) - q_i)^2$. The model receives information about the price only at the final step, where it is used to calculate the loss.

To train the model, I use the Adam optimizer with a learning rate scheduler. With a batch size of 128, running for 5 epochs is sufficient to reach a plateau.

\begin{figure}[!tbh]
    \centering
    \includegraphics[width=\linewidth]{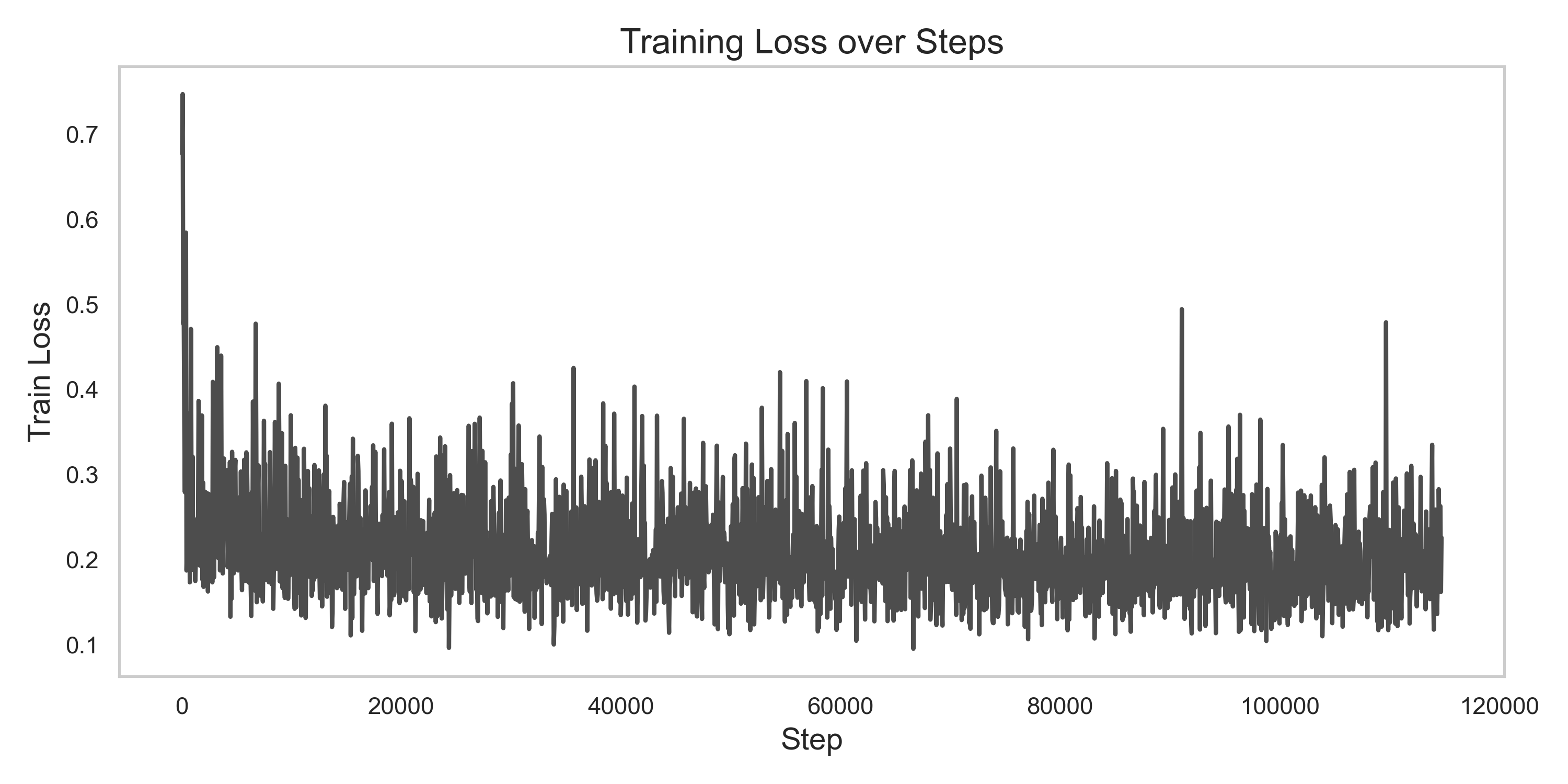}
    \caption{Training loss overs steps}
    \label{fig:enter-label}
\end{figure}

\begin{figure}[!tbh]
    \centering
    \includegraphics[width=\linewidth]{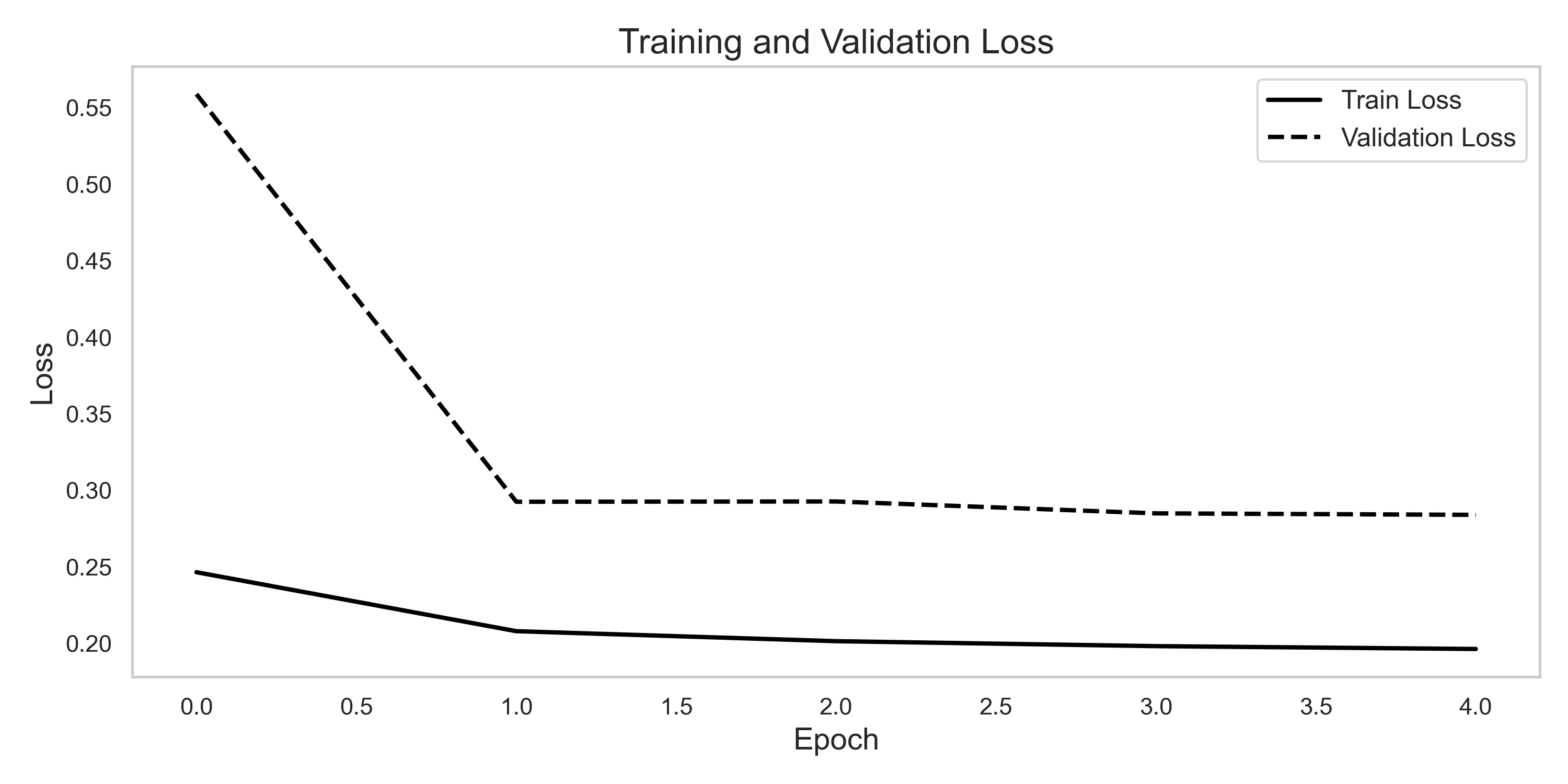}
    \caption{Training and validation loss over epochs}
    \label{fig:enter-label}
\end{figure}

\subsection{Econometric Approach}
Estimation of ~\eqref{eq:econ_model} requires calculating the distances in the product space. The product space is a space of product characteristics that influence the consumers' purchasing decision and their preference for one product over another (substitution drivers). I encode the company's hierarchical product classification and use the observed features (nutrition facts) to form the product space. The analysis results in hundreds of dimensions, which I reduce to 12 dimensions through matrix factorization.\footnote{I also tried autoencoders but the results do not differ significantly.} Finally, I get a 12-dimensional space and calculate the Euclidean distances.  

Unlike the neural network approach, this method requires focusing on a single category. I choose a relatively popular baby food category, which has a wide range of products and is not highly affected by seasonal demand shocks.

\begin{table}
\caption{Descriptive Statistics}
\label{tab:descriptive_statistics}
\scriptsize
\begin{tabular}{lrrrrr}
\toprule
 & Mean & Std. Dev. & Perc. 10\% &  Perc. 90\% \\
\midrule
Log Price & 6.78 & 0.54 & 6.21  & 7.60 \\
CV & 0.16 & 0.11 & 0.07  & 0.26 \\
Log Sales & 7.36 & 0.37 & 6.97  & 7.69 \\
\bottomrule
\end{tabular}
\parbox{0.9\linewidth}{\footnotesize \textit{Note:} This statistic pertains to the baby food category, where the coefficient of variation (CV) is calculated at the product level.}
\end{table}

\begin{figure}[!tbh]
    \centering
    \includegraphics[width=0.67\linewidth]{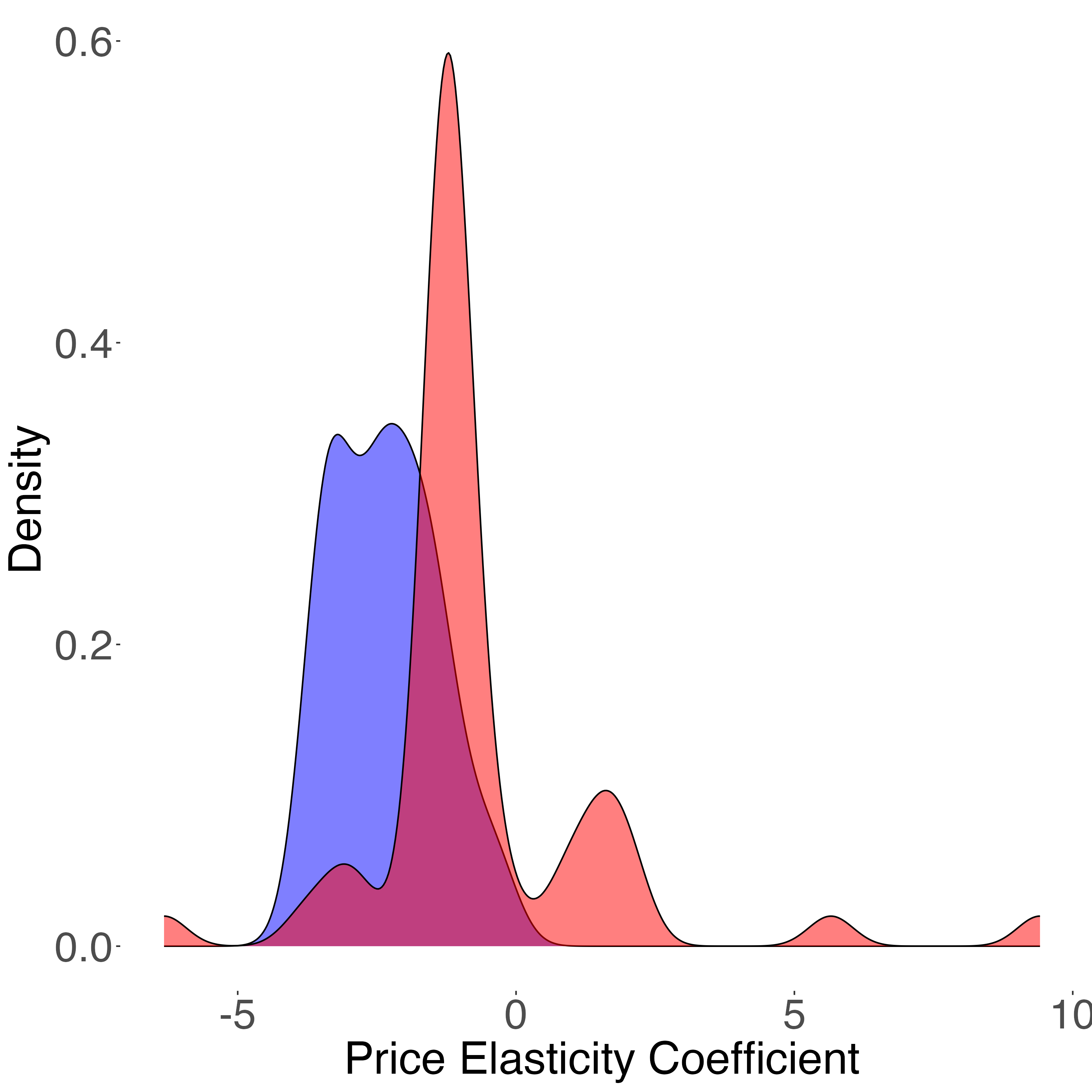}
    \caption{The density plots of elasticity estimates distribution. The red reflects the regression approach while the blue plot comes from the machine learning model output.}
    \label{fig:baby-density}
\end{figure}

Comparing the estimated elasticities from both models,  I find that all elasticity estimates derived from the machine learning model are negative, representing a more meaningful and densely concentrated distribution. In turn, 20\% of estimates from the regression approach are positive. Comparing the results with the simulation, we can say that despite the large amount of data, a small price variation does not allow the regression approach to achieve ``good'' estimates. In Table ~\ref{tab:descriptive_statistics}, the average coefficient of variation for the baby food category is 0.16, which aligns with the simulation data where $\varepsilon = 0.01$ (representing a 1 percent chance that the price will change on a given day). Figure ~\ref{fig:baby-density} represents the distributions of elasticity estimates resulted from the machine learning and regression approaches.

\section{Conclusion} \label{sec:conclusion}
The machine learning model has outperformed traditional econometric models in both simulation and empirical applications. It consistently produces reasonable elasticity estimates, and its fit quality is significantly better than that of other models. The model performs particularly well with limited price variation and does not require specific model specifications, making it easier to evaluate. Further research directions consist of investigating statistical properties of the model's estimates (convergence and asymptotic normality) and comparing it with other models such as the BLP model (Berry, Levinsohn, \& Pakes, 1995), to further assess its performance. It is also important to evaluate the model's performance against experimental methods.

\section{References} \label{references}

Aparicio, D., Metzman, Z., \& Rigobon, R. (2024). The pricing strategies of online grocery retailers. Quantitative Marketing and Economics, 22(1), 1–21.\par
 Berry, S., Levinsohn, J., \& Pakes, A. (1995). Automobile prices in market equilibrium. Econometrica, 63(4), 841–890.\par
 Bertsimas, D., \& Perakis, G. (2006). Dynamic pricing: A learning approach. In Mathematical and computational models for congestion charging (pp. 45–79). Springer.\par
 Besbes, O., \& Zeevi, A. (2009). Dynamic pricing without knowing the demand function: Risk bounds and near-optimal algorithms. Operations Research, 57(6), 1407–1420.\par
 Conlon, C., \& Gortmaker, J. (2020). Best practices for differentiated products demand estimation with PyBLP. The RAND Journal of Economics, 51(4), 1108–1161.\par
 den Boer, A. V. (2015). Dynamic pricing and learning: Historical origins, current research, and new directions. Surveys in Operations Research and Management Science, 20(1), 1–18.\par
 Farias, V. F., \& Van Roy, B. (2010). Dynamic pricing with a prior on market response. Operations Research, 58(1), 16–29.\par
 Farrell, M. H., Liang, T., \& Misra, S. (2021). Deep neural networks for estimation and inference. Econometrica, 89(1), 181–213.\par
 Fisher, M., Gallino, S., \& Li, J. (2017). Competition-based dynamic pricing in online retailing: A methodology validated with field experiments. Management Science, 64(6), 2496–2514.\par
 Harrison, J. M., Keskin, N. B., \& Zeevi, A. (2012). Bayesian dynamic pricing policies: Learning and earning under a binary prior distribution. Management Science, 58(3), 570–586.\par
 Liu, J., Zhang, Y., Wang, X., Deng, Y., \& Wu, X. (2019). Dynamic pricing on e-commerce platform with deep reinforcement learning: A field experiment. arXiv. Preprint.\par
 Nevo, A. (2001). Measuring market power in the ready-to-eat cereal industry. Econometrica, 69(2), 307–342.\par
 Pinkse, J., Slade, M. E., \& Brett, C. (2002). Spatial price competition: A semiparametric approach. Econometrica, 70(3), 1111–1153.\par
 Talluri, K. T., \& van Ryzin, G. J. (2006). The theory and practice of revenue management (Vol. 68). Springer.\par
 Wang, Z., Deng, S., \& Ye, Y. (2014). Close the gaps: A learning-while-doing algorithm for single-product revenue management problems. Operations Research, 62(2), 318–331.\par
 Özer, Ö., \& Phillips, R. (Eds.). (2012). The Oxford handbook of pricing management. Oxford University Press.

\end{document}